\def\rfr#1{eq. (\ref{#1})}

\def\Rfr#1{Eq. (\ref{#1})}

\def\dertt#1#2{\frac{{{d^2}}{#1}}{{{d}}{#2^2}}}

\def\dertt#1#2{\frac{{{d^2}}{#1}}{{{d}}{#2^2}}}

\def\bar{\begin{eqnarray}}
\def\ear{\end{eqnarray}}
\def\bb{\bibitem}
\def\eqi{\begin{equation}}
\def\eqf{\end{equation}}
\def\eqia{\begin{eqnarray}}
\def\eqfa{\end{eqnarray}}
\def\rp#1#2{\frac{#1}{#2}}

\def\lb#1{\label{#1}}

\def\bds#1{\boldsymbol{#1}}

\documentclass[mathleft]{an}
\usepackage{graphicx}
\usepackage{times}
\usepackage{amsmath}
\begin{document}

% The following seven commands are intended for editorial usage and should be ignored by
% the author(s).
\Pagespan{861}{}% Document's page range.
% If second parameter is left empty, the last page is computed automatically.
\Yearpublication{2009}%
\Yearsubmission{2009}%
%\Month{8}%
\Volume{330}%
\Issue{8}%
 \DOI{10.1002/asna.200911239}%

\title{Galactic   Sun's   motion in the Cold Dark Matter, MOdified Newtonian Dynamics and MOdified Gravity scenarios}

\author{L. Iorio\inst{1}\fnmsep\thanks{Corresponding author:
  \email{lorenzo.iorio@libero.it}\newline}
%Example
%for footnote, note the usage of the \texttt{fnmsep}
%command as separator between institute number and footnote mark}
%\and  G.H. Ostwriter\inst{2,3}
}
\titlerunning{Galactic Sun's motion in CDM, MOND and MOG}
\authorrunning{L. Iorio}
\institute{
INFN-Sezione di Pisa, Viale Unit\`{a} di Italia 68,
I-70125 Bari, Italy
}

%\received{30 May 2005}
%\accepted{11 Nov 2005}
%\publonline{later}

\keywords{gravitation -- Galaxy: solar neighborhood -- cosmology: dark matter -- celestial mechanics, stellar dynamics}

\abstract{ %
  We numerically integrate  the equations of motion of the Sun in Galactocentric Cartesian rectangular coordinates for $-4.5$ Gyr $\leq t\leq 0$ in Newtonian mechanics with two different models for the Cold Dark Matter (CDM) halo,  in MOdified Newtonian Dynamics (MOND) and  in MOdified Gravity (MOG) without resorting to CDM. The initial conditions used come from the latest kinematical determination of the 3D  Sun's motion in the Milky Way (MW) by assuming for the rotation speed of the Local Standard of Rest (LSR) the recent value $\Theta_0=268$ km s$^{-1}$ and the IAU recommended value $\Theta_0=220$ km s$^{-1}$; the Sun is assumed located at 8.5 kpc from the Galactic Center (GC).
  For   $\Theta_0=268$ km s$^{-1}$ the birth of the Sun, 4.5 Gyr ago, would have occurred at large Galactocentric distances ($12-27$ kpc depending on the model used), while for  $\Theta_0=220$ km s$^{-1}$  it would have occurred at about $8.8-9.3$ kpc for almost all the models used. The integrated trajectories are far from being circular, especially for $\Theta_0=268$ km s$^{-1}$, and differ each other with the CDM models yielding the widest spatial extensions for the Sun's orbital path.  }

\maketitle
\section{Introduction}
In several astrophysical systems like, e.g., spiral galaxies and clusters of galaxies a discrepancy between the observed kinematics of their exterior parts and the predicted one on the basis of the Newtonian dynamics and the matter detected from the emitted electromagnetic radiation (visible stars and gas clouds) was present  since the pioneering studies by \cite{Zwi} (he postulated the existence of undetected, baryonic matter; today, it is believed that the hidden mass is constituted by non-baryonic, weakly interacting particles) on the Coma cluster, and by \cite{Bos} and \cite{Rub} on spiral galaxies. More precisely, such an effect shows up in the galactic  velocity rotation curves (\cite{flatt,flattona}) whose typical pattern after a few kpc from the center differs from the Keplerian $1/\sqrt{r}$ fall-off expected from the usual dynamics applied to the electromagnetically-observed matter.

As a possible solution of this puzzle, the existence of non-baryonic, weakly-interacting Cold Dark (in the sense that its existence is indirectly inferred only from its gravitational action, not from emitted electromagnetic radiation) Matter (CDM) was proposed to reconcile the predictions with the observations (\cite{Rub83}) in the framework of the standard gravitational physics.

Oppositely, it was postulated that the Newtonian laws of gravitation have to be modified on certain acceleration scales to correctly account for the observed anomalous kinematics of such astrophysical systems without resorting to still undetected exotic forms of matter.
One of the most phenomenologically successful modification of the inverse-square Newtonian law, mainly with respect to spiral galaxies, is the MOdified Newtonian Dynamics (MOND)  (\cite{Mil83a,Mil83b,Mil83c}) which, as we will see below, yields a $\approx 1/r$ gravitational force for very small accelerations.

Another alternative theoretical framework recently put forth to also explain, among other astrophysical and cosmological observations, the observed kinematics of the outskirts of galaxies without resorting to CDM is MOdified Gravity (MOG) (\cite{Mof}) which yields a long-range Yukawa-like modification of the Newtonian inverse-square law.

In this paper we want to investigate the motion of the Sun in the Milky Way (MW) over timescales of the order of its lifetime, i.e. 4.5 Gyr, in CDM, MOND and MOG scenarios in view of the recent accurate measurements of the solar kinematical parameters (\cite{Reid}). This may help in discriminating, at least in principle, the different theoretical scenarios examined; moreover, it will be interesting to see if the Galactic orbital motions of the Sun backward in time  predicted by CDM, MOND and MOG are compatible with known and accepted knowledge concerning its formation and the evolution of the life in our planet. This may also yield an independent consistency test of the new kinematical determinations of MW  (\cite{Reid}). Finally, we note that the approach presented here may be used, in principle, also for other dynamical features.
\section{The models used}
\subsection{The CDM NFW model}
The CDM model tested by \cite{Xue} with several  Blue Horizontal-Branch (BHB) halo stars  consists of  three components. Two of them are for the disk
\eqi U_{\rm disk}=-\rp{GM_{\rm disk}\left[1-\exp\left(-\rp{r}{b}\right)\right]}{r},\lb{disk}\eqf where $b$ is the disk scale length,
and the bulge
\eqi U_{\rm bulge} = -\rp{GM_{\rm bulge}}{r+c_0},\lb{bulge}\eqf
where $c$ is the bulge scale radius.
The third component is for the CDM NFW halo (\cite{NFW})
\eqi U_{\rm NFW}=-\rp{4\pi G\rho_s r_{\rm vir}^3}{c^3 r}\ln\left(1+\rp{cr}{r_{\rm vir}}\right),\lb{halo}\eqf
with  $r_{\rm vir}$ is the radius parameter  and
\eqi \rho_s=\rp{\rho_{\rm cr}\Omega_{\rm m}\delta_{\rm th}}{3}\rp{c^3}{\ln\left(1+c\right)-\rp{c}{1+c}},\eqf
in which $\Omega_{\rm m}$ is the fraction of matter (including baryons and DM) to the critical density, $\delta_{\rm th}$ is critical overdensity of the virialized system, $c$ is the concentration parameter, and
\eqi \rho_{\rm cr}=\rp{3 H_0^2}{8\pi G} \eqf is the critical density of the Universe determined by the Hubble parameter $H_0$ at redshift $0$.
The values used for the parameters entering \rfr{disk} and \rfr{bulge} are in Table \ref{buldis};
\begin{table}
% \centering%%%
\caption{Parameters of the disk and bulge models by \cite{Xue}.}
\label{buldis}
\begin{tabular}{cccc}\hline
$M_{\rm disk}$ & $M_{\rm bulge}$  & $b$  & $c_0$ \\
(M$_{\odot}$) & (M$_{\odot}$)  & (kpc)  & (kpc) \\
\hline
 $5\times 10^{10}$   & $1.5\times 10^{10}$  & $4$  & $0.6$\\
\hline
\end{tabular}
\end{table}
those entering \rfr{halo} are in Table \ref{NFW}.
\begin{table}
% \centering%%%
\caption{Parameters of the CDM NFW halo model by \cite{Xue}. The values quoted for $r_{\rm vir}$ and $c$ come from an average of those by \cite{Xue}.}
\label{NFW}
\begin{tabular}{ccccc}\hline
$\Omega_{\rm m}$ & $\delta_{\rm th}$  & $r_{\rm vir}$  & $c$ & $H_0$\\
- & -  & (kpc)  & (kpc) & (km s$^{-1}$ Mpc$^{-1}$)\\
\hline
 $0.3$   & $340$  & $273.2$  & $8.9$  & 65\\
\hline
\end{tabular}
\end{table}
\subsection{The logarithmic CDM halo}
Another model widely used
consists of the \cite{Miy75} disk
\eqi U_{\rm disk} = -\rp{\xi GM_{\rm disk}}{\sqrt{x^2+y^2+\left(k+\sqrt{z^2+{B}^2}\right)^2}}, \lb{MiyaNaga}\eqf
where $k$ and $B$ are the disk scale length and the disk scale height, respectively,
the \cite{Plum11} bulge
\eqi U_{\rm bulge} = -\rp{GM_{\rm bulge}}{r+C}, \lb{Plumm}\eqf
where $C$ is the bulge scale radius,
and the logarithmic CDM halo   (\cite{BinTrem87})
\eqi U_{\rm halo} = v^2_{\rm halo}\ln\left(r^2+d^2\right),\ ({\rm spherical\ halo})\lb{BinTrem}\eqf
where $d$ is the DM halo scale length and $v_{\rm halo}$ describes the DM halo dispersion speed (which is related to the total DM halo mass).
The parameters's values are in Table \ref{Mya}.
\begin{table}
% \centering%%%
\caption{Parameters of the logarithmic CDM halo model; values by \cite{Law}.}
\label{Mya}
\begin{tabular}{cccccc}\hline
$\xi$ & $k$  & $B$ & $C$  & $v_{\rm halo}$ & $d$\\
 - & (kpc)  & (kpc)  & (kpc) & km s$^{-1}$ & (kpc)\\
\hline
 $1$ & $6.5$ & $0.26$ & $0.7$ & $114$ & $12$\\
\hline
\end{tabular}
\end{table}
The masses of the disk and the bulge used by \cite{Law} are those by
\cite{John99}: $M_{\rm disk}=1\times 10^{11}$ M$_{\odot}$, $M_{\rm bulge}=3.4\times 10^{10}$ M$_{\odot}$  yielding a total baryonic mass of $M=1.34\times 10^{11}$ M$_{\odot}$; however, such a value is almost twice the most recent estimate ($M=6.5\times 10^{10}$ M$_{\odot}$) by \cite{McG} who includes the gas mass as well and yield $M_{\rm disk}=2.89\times 10^{10}$ M$_{\odot}$ and $M_{\rm bulge}=2.07\times 10^{10}$ M$_{\odot}$.   The model of \rfr{MiyaNaga}-\rfr{BinTrem}, with the parameters' values by \cite{Law} and \cite{John99}, has been recently used by \cite{Will09}
to study the motion of the \cite{Gril} tidal stellar stream at galactocentric distance of $r\approx 16-18$ kpc; \cite{Read} used it to study the motion of the tidal debris of the Sagittarius dwarf at 17.4 kpc from the center of MW.  More specifically, the CDM halo model of \rfr{BinTrem} corresponds to a CDM halo mass
\eqi M_{\rm halo}=\rp{2 v_{\rm halo}^2 r^3}{G(r^2+d^2)},\eqf
so that
\eqi M_{\rm halo}(r=60\ {\rm kpc})= 3.5\times 10^{11}\ {\rm M}_{\odot},\eqf
in agreement with the value by \cite{Xue}
\eqi  M_{\rm halo}(r=60\ {\rm kpc})= (4.0\pm 0.7)\times 10^{11}\ {\rm M}_{\odot}.\eqf
\subsection{MOdified Newtonian Dynamics}
MOND  postulates that
for systems  experiencing total gravitational acceleration  $A < A_0$, with (\cite{Bege}) \eqi A_0= (1.2\pm 0.27)\times 10^{-10} \ {\rm m\ s}^{-2},\eqf
\eqi A\rightarrow A_{\rm MOND}=-\rp{\sqrt{A_0GM}}{r}.\lb{MOND}\eqf   More precisely, it holds \eqi A = \rp{A_{\rm Newton}}{\mu(X)},\ X\equiv\rp{A}{A_0};\lb{appromond}\eqf
$\mu(X)\rightarrow 1$ for $X\gg 1$, i.e. for large accelerations (with respect to $A_0$), while $\mu(X)\rightarrow X$ yielding \rfr{MOND} for $X\ll 1$, i.e. for small accelerations.
Concerning the interpolating function $\mu(X)$, it recently turned out that the simple form
(\cite{Fam})
\eqi \mu(X) = \rp{X}{1+X}.\lb{mu1}\eqf
yields very good results in fitting the terminal velocity curve of MW, the rotation curve of the standard external galaxy NGC 3198 (\cite{Fam, kazzo, scassa}) and of a sample of 17 high
surface brightness, early-type disc galaxies (\cite{Noor});
\rfr{appromond} becomes
\eqi A = \rp{A_{\rm Newton}}{2}\left(1+\sqrt{1+\rp{4 A_0}{A_{\rm Newton}}}\right)\lb{oleee}\eqf
with \rfr{mu1}.
Thus, in the following we will use \rfr{oleee}.
\subsection{MOdified Gravity}
MOG is a fully covariant
theory of gravity which is based on the existence of a massive vector field
coupled universally to matter. The theory yields a Yukawa-like modification of
gravity with three constants which, in the most general case, are running; they
are present in the theory's action as scalar fields which represent the
gravitational constant, the vector field coupling constant, and the vector
field mass.   An approximate solution of the
MOG field equations by \cite{cazMOF} allows to compute their values as functions of the
source's mass. The resulting Yukawa-type modification of the
inverse-square Newton's law in the gravitational field of a central mass $M$ is
\eqi A = -\rp{G_{\rm N}M}{r^2}\left\{1+\alpha\left[1-\left(1+\mu r\right)\exp\left(-\mu r\right)\right]\right\},\lb{Mog}\eqf
with
\begin{eqnarray}
  \alpha &\simeq & \rp{M}{\left(\sqrt{M}+C^{'}_1\right)^2}\left(\rp{G_{\infty}}{G_{\rm N}}-1\right), \\
  \mu &\simeq & \rp{C_2^{'}}{\sqrt{M}},
\end{eqnarray}
where  $G_{\rm N}$ is the Newtonian gravitational constant  and
\begin{eqnarray}
  G_{\infty} &\simeq & 20 G_{\rm N}, \\
  C_1^{'} &\simeq & 25,000\sqrt{{\rm M}_{\odot}}, \\
  C_2^{'} &\simeq & 6,250\sqrt{{\rm M}_{\odot}}\ {\rm kpc}^{-1}.
\end{eqnarray}
Such values have been obtained by \cite{cazMOF} as a result of the fit of the velocity rotation curves of some
galaxies in the framework of the searches for an explanation of the rotation curves of galaxies without resorting to CDM.
The validity of \rfr{Mog} in the Solar System has been recently questioned by \cite{IorMOG}.

For $M=6.5\times 10^{10}$ M$_{\odot}$ we have the values of Table \ref{MOOOG}.
 \begin{table}
% \centering%%%
\caption{MOG parameters $\alpha$ and $\lambda$ for $M=6.5\times 10^{10}$ M$_{\odot}$. }
\label{MOOOG}
\begin{tabular}{cc}\hline
$\alpha$ & $\lambda$\\
-& (kpc)\\
 \hline
 $16$  & $41$ \\
\hline
\end{tabular}
\end{table}
\section{Solar motions backward in time}
We will numerically integrate with\footnote{We will use the default options of NDSolve.} MATHEMATICA the Sun's equations of motion in Cartesian rectangular coordinates for $-4.5$ Gyr $\leq t\leq 0$ wit the initial conditions of Table \ref{statevec}; in it we use the recently estimated value $\Theta_0=268$ km s$^{-1}$ for the rotation speed of the Local Standard of Rest (LSR) (\cite{Reid}).
\begin{table}
% \centering%%%
\caption{Galactocentric initial conditions for the Sun (\cite{Reid}); the positive $y$ axis is directed from the Galactic Center (GC) to the Sun, the positive $x$ axis is directed toward the Galactic rotation, the positive $z$ axis is directed toward the  North Galactic Pole (NGP). We have used (\cite{Reid}) $\Theta_0 = 268$ km s$^{-1}$ for the rotation speed of LSR, and $U_0=10.3$ km s$^{-1}$, $V_0=15.3$ km s$^{-1}$, $W_0=7.7$ km s$^{-1}$ for the standard solar motions toward GC, $\ell=90$ deg and NGP, respectively; thus, with our conventions, $\dot x_0=V_0+\Theta_0,\ \dot y_0 = -U_0,\ \dot z_0 =W_0$. See Fig. 7 of (\cite{Reid}).}
\label{statevec}
\begin{tabular}{ccccccc}\hline
$x_0$ & $y_0$  & $z_0$  & $\dot x_0$ & $\dot y_0$ & $\dot z_0$\\
(kpc) & (kpc)  & (kpc)  & (km s$^{-1}$) & (km s$^{-1}$) & (km s$^{-1}$) \\
\hline
 $0$  & $8.5$  & $0.02$  & $15.3+268$  & $-10.3$  & $7.7$  \\
\hline
\end{tabular}
\end{table}

We will use  both the NFW and logarithmic halo models for CDM which have been tested independently of the solar motion itself. In particular, we will numerically integrate the three scalar differential equations, written in cartesian coordinates, corresponding to the vector differential equation
\eqi \dertt{\bds r} t=-\bds\nabla U,\eqf where $U$ is the sum of \rfr{disk}-\rfr{halo} with the values of Table \ref{NFW} in the first case (NFW), and of \rfr{MiyaNaga}-\rfr{BinTrem} with the values of Table \ref{Mya} in the second case (logarithmic); we use the figures of Table \ref{buldis} in both cases. For the two models of modified gravity considered we will put the cartesian components of \rfr{oleee} (MOND) and \rfr{Mog} (MOG) in the right-hand-sides of the differential equations to be integrated. Concerning MOND, let us note that
\Rfr{appromond}, from which \rfr{oleee} has been derived by means of \rfr{mu1} for $\mu(X)$, strictly holds for co-planar, spherically and axially symmetric mass distributions; otherwise, the full  modified (non-relativistic) Poisson equation  (\cite{joint})
\eqi \bds\nabla\bds\cdot\left[\mu\left(\rp{|\bds\nabla U|}{A_0}\right)\bds\nabla U\right]=4\pi G\rho\eqf
must be used. However, in this case, the baryonic mass distributions of the bulge and the disk (see \rfr{disk}-\rfr{bulge} and \rfr{MiyaNaga}-\rfr{Plumm})  just satisfy the symmetry conditions that allow to use \rfr{appromond}.
Moreover, $A_{\rm N}/A_0\approx 1$ for $M=6,5\times 10^{10}$ M$_{\odot}$. This allows to neglect the so-called External Field Effect (EFE) (\cite{San02,Mil08}) because it should amount to about $A_{\rm e}/A_0=0.01-0.02$ for MW (\cite{WuMag}). Indeed, it can be shown that, for \rfr{mu1}, the MOND acceleration, including EFE, is
{
\setlength{\mathindent}{0pt}%%%
 \footnotesize{\eqi A=\rp{A_{\rm N}}{2}\left[1-\rp{A_{\rm e}}{A_{\rm N}} + \sqrt{\left(1-\rp{A_{\rm e}}{A_{\rm N}}\right)^2
+\rp{4A_0}{A_{\rm N}}\left(1+\rp{A_{\rm e}}{A_0}\right)}
\right],\lb{mega}\eqf }
}
which just reduces  to \rfr{oleee} when $A_{\rm e}\ll A_{\rm N},\ A_{\rm e}\ll A_0$, as in this case.

\begin{figure}
\includegraphics[width=\columnwidth]{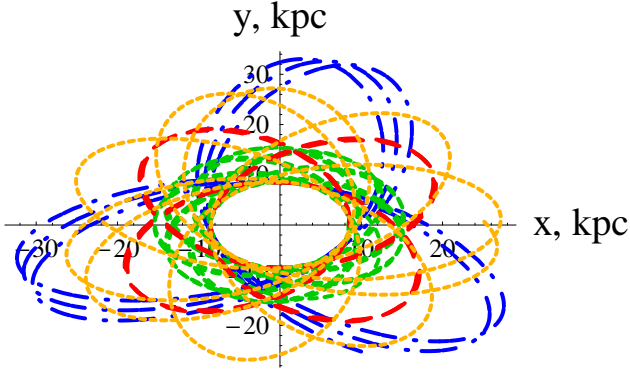}
%[width=50mm,height=28mm]
\caption{Section in the $\{x,y\}$ plane of the numerically integrated trajectories of the Sun for a) CDM logarithmic halo (dash-dotted blue line) b) CDM NFW halo (dotted yellow line) c) MOND with $\mu=X/(1+X)$ (dashed green line) c) MOG (dashed red line). The time span is $-4.5$ Gyr $\leq t\leq 0$. The initial conditions are those of Table \ref{statevec} with (\cite{Reid}) $\Theta_0=268$ km s$^{-1}$.}
\label{Sun_xy}
\end{figure}

\begin{figure}
\includegraphics[width=\columnwidth]{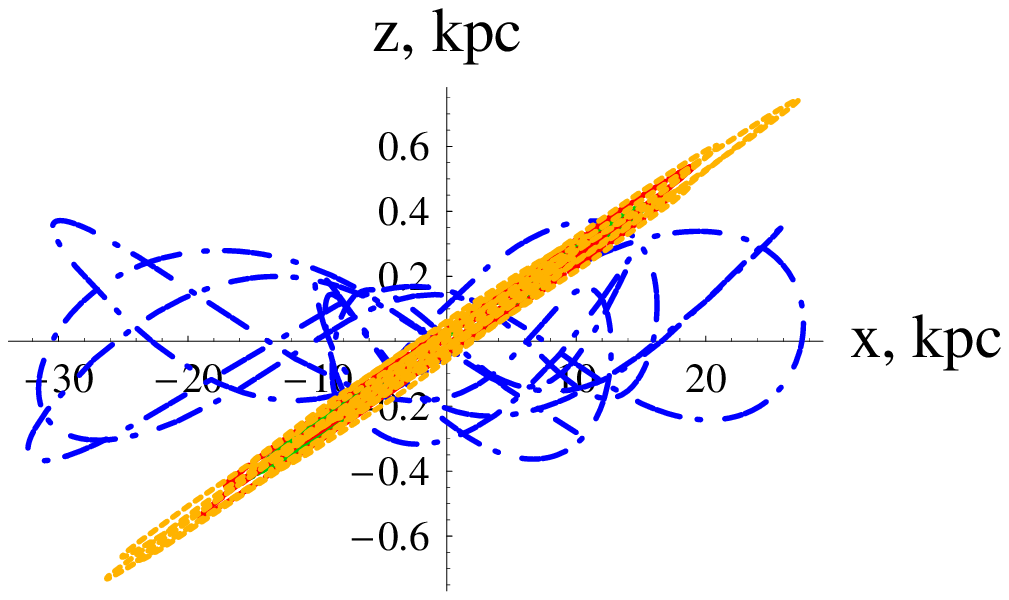}
%[width=50mm,height=28mm]
\caption{Section in the $\{x,z\}$ plane of the numerically integrated trajectories of the Sun for a) CDM logarithmic halo (dash-dotted blue line) b) CDM NFW halo (dotted yellow line) c) MOND with $\mu=X/(1+X)$ (dashed green line) c) MOG (dashed red line). The time span is $-4.5$ Gyr $\leq t\leq 0$. The initial conditions are those of Table \ref{statevec} with (\cite{Reid}) $\Theta_0=268$ km s$^{-1}$.}
\label{Sun_xz}
\end{figure}

\begin{figure}
\includegraphics[width=\columnwidth]{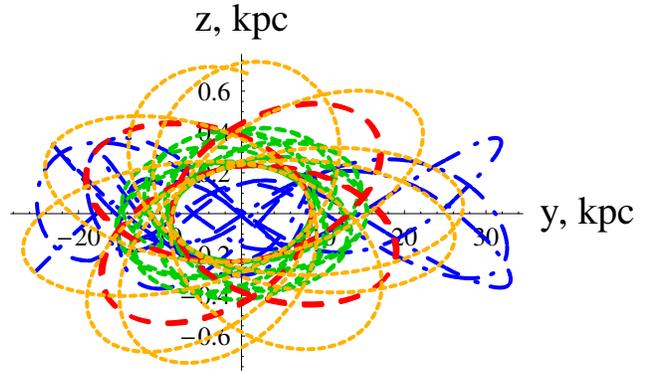}
%[width=50mm,height=28mm]{Sun_yz.eps}
\caption{Section in the $\{y,z\}$ plane of the numerically integrated trajectories of the Sun for a) CDM logarithmic halo (dash-dotted blue line) b) CDM NFW halo (dotted yellow line) c) MOND with $\mu=X/(1+X)$ (dashed green line) c) MOG (dashed red line). The time span is $-4.5$ Gyr $\leq t\leq 0$. The initial conditions are those of Table \ref{statevec} with (\cite{Reid}) $\Theta_0=268$ km s$^{-1}$.}
\label{Sun_yz}
\end{figure}

In Figure \ref{Sun_xy}-Figure \ref{Sun_yz} we plot the orbital sections in the Galactocentric coordinate planes of the numerically integrated trajectories of the Sun from now to 4.5 Gyr ago.
The orbital trajectories of the Sun are quite different in CDM, MOND and MOG; the largest spatial extension occur for the CDM models, especially for the one characterized by the CDM logarithmic halo
The Galactocentric coordinates and distances of the Sun at $t=-4.5$ Gyr are quoted in Table \ref{tavola2}.
\begin{table}
% \centering%%%
\caption{Galactocentric coordinates and distance of the Sun at $t=-4.5$ Gyr according to our numerical integrations of the equations of motion in CDM (logarithmic halo), CDM (NFW halo), MOND  and MOG for (\cite{Reid}) $\Theta_0=268$ km s$^{-2}$. The initial conditions of Table \ref{statevec} have been used.}
\label{tavola2}
\begin{tabular}{ccccc}\hline
Model  & $x$ & $y$  & $z$  & $r$ \\
 & (kpc) & (kpc) & (kpc) & (kpc) \\
\hline
CDM log & $9.9$ & $-25.2$ & $-0.3$ & $27.1$ \\
CDM NFW & $25.1$ & $0.7$  & $0.7$ & $25.1$\\
MOND $\left(\mu = \rp{X}{1+X}\right)$ & $4.4$  & $11.1$  & $0.1$  & $12.0$\\
MOG & $15.4$ & $-3$  & $0.4$ & $15.7$\\
\hline
\end{tabular}
\end{table}
It can be noted that the models considered place the birth of the Sun from a minimum of 12.0 kpc (MOND) to a maximum of $25-27$ kpc (CDM) from GC.
This may pose problems concerning the birth of the Earth itself and the development of complex life on it because of, e.g., the presumable low level of
metallicity (the amount of elements heavier than hydrogen and helium) in so distant regions. For a discussion and the implications of the concept of Galactic Habitable Zone (GHZ), see \cite{GHZ1}, \cite{Gonz} and \cite{GHZ2}.   In particular, see Figure 5 by \cite{GHZ2} which shows that the probability of having Earths formed 4 and 8 Gyr after the formation of MW, which roughly corresponds to the birth of the Sun by assuming a Milky Way age of about $10-12$ Gyr, is practically null at Galactocentric distances larger than $10-15$ kpc.

\begin{figure}
\includegraphics[width=\columnwidth]{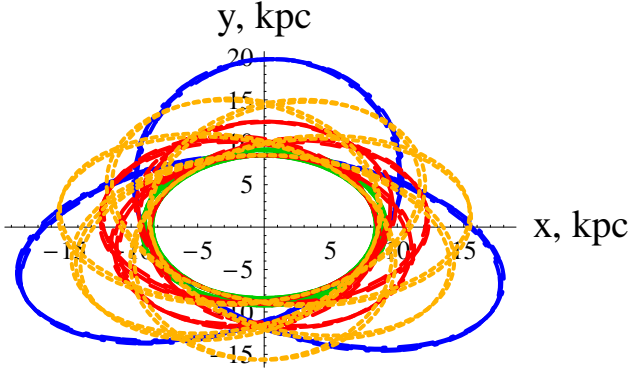}
%[width=50mm,height=28mm]
\caption{Section in the $\{x,y\}$ plane of the numerically integrated trajectories of the Sun for a) CDM logarithmic halo (dash-dotted blue line) b) CDM NFW halo (dotted yellow line) c) MOND with $\mu=X/(1+X)$ (dashed green line) c) MOG (dashed red line). The time span is $-4.5$ Gyr $\leq t\leq 0$. The initial conditions are those of Table \ref{statevec} with the IAU recommended value $\Theta_0=220$ km s$^{-1}$.}
\label{Sun_xy_220}
\end{figure}

\begin{figure}
\includegraphics[width=\columnwidth]{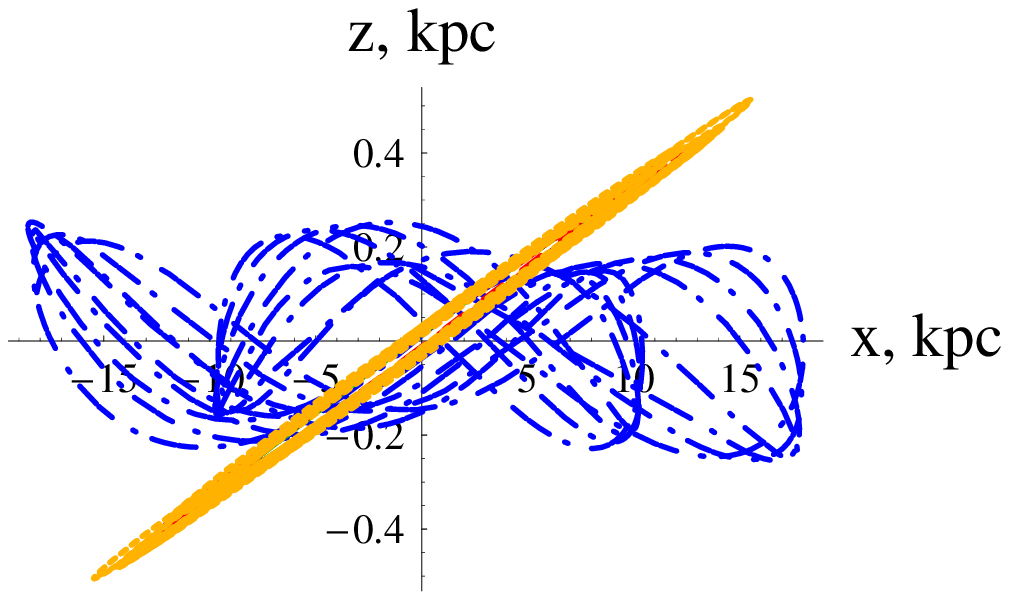}
%[width=50mm,height=28mm]
\caption{Section in the $\{x,z\}$ plane of the numerically integrated trajectories of the Sun for a) CDM logarithmic halo (dash-dotted blue line) b) CDM NFW halo (dotted yellow line) c) MOND with $\mu=X/(1+X)$ (dashed green line) c) MOG (dashed red line). The time span is $-4.5$ Gyr $\leq t\leq 0$. The initial conditions are those of Table \ref{statevec} with the IAU recommended value $\Theta_0=220$ km s$^{-1}$.}
\label{Sun_xz_220}
\end{figure}

\begin{figure}
\includegraphics[width=\columnwidth]{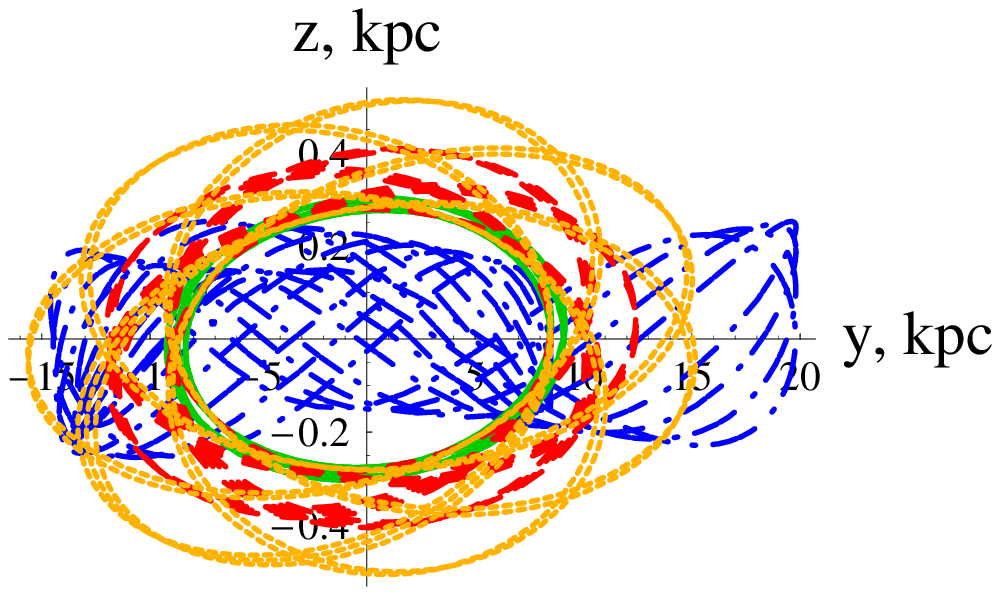}
%[width=50mm,height=28mm]{Sun_yz.eps}
\caption{Section in the $\{y,z\}$ plane of the numerically integrated trajectories of the Sun for a) CDM logarithmic halo (dash-dotted blue line) b) CDM NFW halo (dotted yellow line) c) MOND with $\mu=X/(1+X)$ (dashed green line) c) MOG (dashed red line). The time span is $-4.5$ Gyr $\leq t\leq 0$. The initial conditions are those of Table \ref{statevec} with the IAU recommended value $\Theta_0=220$ km s$^{-1}$.}
\label{Sun_yz_220}
\end{figure}

In Figure \ref{Sun_xy_220}-Figure \ref{Sun_yz_220} we repeat the same integrations by using the IAU recommended value $\Theta_0=220$ km s$^{-1}$.
The Galactocentric coordinates and distances of the Sun at $t=-4.5$ Gyr for $\Theta_0=220$ km s$^{-1}$ are quoted in Table \ref{tavola3}.
\begin{table}
% \centering%%%
\caption{Galactocentric coordinates and distance of the Sun at $t=-4.5$ Gyr according to our numerical integrations of the equations of motion in CDM (logarithmic halo), CDM (NFW halo), MOND  and MOG. The initial conditions of Table \ref{statevec} have been used with the IAU recommended value $\Theta_0=220$ km s$^{-2}$.}
\label{tavola3}
\begin{tabular}{ccccc}\hline
Model  & $x$ & $y$  & $z$  & $r$ \\
 & (kpc) & (kpc) & (kpc) & (kpc) \\
\hline
CDM log & $-18.1$ & $-4.7$ & $0.1$ & $18.8$ \\
CDM NFW & $7.1$ & $-5.2$  & $0.2$ & $8.8$\\
MOND $\left(\mu = \rp{X}{1+X}\right)$ & $6.1$  & $6.5$  & $0.2$  & $8.9$\\
MOG & $9.2$ & $1.1$  & $0.3$ & $9.3$\\
\hline
\end{tabular}
\end{table}
Now the situation is quite different because, apart from the logarithmic CDM halo, all the other models locate the birth of the Sun at $8.8-9.3$ kpc from GC

\section{Summary and Conclusions}
By using the latest kinematical determinations of the full 3D solar motion in the Milky Way, implying a LSR rotation of $\Theta_0=268$ km s$^{-1}$, we used them as initial conditions for numerically integrating the equations of motion of the Sun backward in time ($-4.5$ Gyr $\leq t\leq 0$) in Newtonian mechanics with two models for the CDM halo, in MOND and in MOG. As a result, the orbital trajectories are not circular and differ each other, with the CDM  models yielding the widest spatial extensions of the Sun's orbit. It turns out that for $t=-4.5$ Gyr the Sun is at quite large Galactocentric distances ($r\approx 12-27$ kpc depending on the models). Instead, by using the standard IAU value $\Theta_0=220$ km s$^{-1}$ the situation is different: the orbits are less wide and at $t=-4.5$ Gyr the Sun is at $8.8-9.3$ kpc from GC for almost all the models considered.

\acknowledgements
I thank M. Masi for useful references and discussions.
%We appreciate the cooperation with the authors of AN.
%Based on photographic data obtained using The UK Schmidt Telescope.
%  The UK Schmidt Telescope was operated by the Royal Observatory
%  Edinburgh, with funding from the UK Science and Engineering Research
%  Council, until 1988 June, and\linebreak%%%%%
%  thereafter by the Anglo-Australian
%  Observatory.  Original plate material is copyright (c) the Royal
%  Observatory Edinburgh and the Anglo-Australian Observatory.  The
%  plates were processed into the present compressed digital form with
%  their permission.  The Digitized Sky Survey was produced at the Space
%  Telescope Science Institute under US Government grant NAG W-2166.

\newpage%%%%%%%%%%%%%%%%%%%%%%%%%%%%%%%%%%%%%%%%%%%%%%%%%%%%%%

%\appendix
%
%\section{This is the title of the first appendix}
%Larger tables, collections of images, spectra or similar kind of data shall be
%presented in the appendix section rather than in the main body of the
%text. Several appendices can be separated by the \verb+\section{+{\it title
%of appendix}\verb+}+ command. They are enclosed in the
%\verb+appendix+ environment.

\end{document}